\documentclass[twocolumn]{aastex631}

\usepackage{url}
\usepackage{xcolor}
\usepackage{xspace}
\usepackage[caption=false]{subfig}
\usepackage{mathtools}

\usepackage{threeparttable} 
\let\tablenum\relax
\usepackage{siunitx}

\DeclarePairedDelimiterXPP\BigOSI[2]%
  {\mathcal{O}}{(}{)}{}%
  {\SI{#1}{#2}}

\newcommand{\src}[0]{FRB 20240114A}
\newcommand{\ppcc}[0]{pc\,cm$^{-3}$\xspace}
\newcommand{\dmhost}[0]{DM$_{\rm host}$\xspace}

\begin{document}

\title{Varying activity and the burst properties of \src{} probed with GMRT down to 300 MHz}

%% %%
%% 
\author[0009-0002-0330-9188]{Ajay Kumar}
\email{akumar@ncra.tifr.res.in}
\affiliation{National Centre for Radio Astrophysics (NCRA - TIFR), Pune - 411007, India}

\author[0000-0002-0862-6062]{Yogesh Maan}
\email{ymaan@ncra.tifr.res.in}
\affiliation{National Centre for Radio Astrophysics (NCRA - TIFR), Pune - 411007, India}
\author[0000-0002-5342-163X]{Yash Bhusare}
\email{ybhusare@ncra.tifr.res.in}
\affiliation{National Centre for Radio Astrophysics (NCRA - TIFR), Pune - 411007, India}

\begin{abstract}
Repeating Fast Radio Bursts (FRBs) can exhibit a wide range of burst repetition rates, from none to hundreds of bursts per hour.
Here, we report the detection and characteristics of 60 bursts from the recently discovered FRB 20240114A, observed with the upgraded Giant Metrewave Radio Telescope (uGMRT) in the frequency ranges 300-500\,MHz and 550-750\,MHz. The majority of the bursts show narrow emission-bandwidth with $\Delta\nu/\nu \sim$ 10\%. All of the bursts we detect are faint ($<$10\,Jy\,ms), and thus probe the lower end of the energy distribution. We determine the rate function for FRB 20240114A at 400 MHz, and downward drift rates at 400 and 650\,MHz, and discuss our measurements in the context of the repeating FRB population. We observe sudden variations in the burst activity of FRB 20240114A over time. From our data as well as the publicly available information on other observations of FRB 20240114A so far, there is an indication that FRB 20240114A potentially exhibits chromaticity in its burst activity. While the burst properties of FRB 20240114A are similar to other repeating FRBs, the frequency-dependent activity, if established, could provide crucial clues to the origin of repeating FRBs. We also place the most stringent 5$\sigma$ upper limits of 600\,$\mu$Jy and 89\,$\mu$Jy on any persistent radio source (PRS) associated with FRB 20240114A at 400\,MHz and 650\,MHz, respectively, and compare these with the luminosity of the known PRSs associated with FRB121102A and FRB190520B. 
\end{abstract}


\section{Introduction} \label{sec:intro}
Fast radio bursts (FRB) are coherent bright radio flashes of cosmological origin with duration ranging from a few micro to several milliseconds. Their exceptional high dispersion measure (DM) along the line of sight can not be explained by the Galactic electron content, thus predicting their extragalactic origin. The extragalactic nature is now confirmed with several repeating FRBs localized to a variety of galaxies \citep{2017Natur.541...58C, 2019Sci...365..565B, 2020Natur.577..190M, 2020ApJ...895L..37B, 2020ApJ...903..152H}. The origin and emission mechanism of FRBs still remain open questions. Currently there are $\sim $800 published FRBs, of which some are repeating in nature. Discovery of the first repeating FRB, FRB 20121102A \citep{2014ApJ...790..101S}, ruled out cataclysmic events as the source for repeating FRBs. Currently there are close to sixty repeaters known \footnote{\url{https://www.chime-frb.ca/repeaters}} but only a few are known to be extremely active occasionally. 
A few repeating FRBs show periodicity in their activity, e.g., FRB 180916 exhibits a periodicity of 16.35 days while FRB 121102A shows a tentative periodicity of 157 days \citep{2020Natur.582..351C,Rajwade_2020}. Some FRBs also show very high bursts rates ($>$100 bursts/hr) occasionally, e.g., FRB201124A and FRB220912A \citep{fast_r67,R117_gbt}. Several repeaters have been followed up or monitored over a long duration of time and their properties very well studied but we still we do not have source models that can explain all their observed properties. Discovery of highly active repeaters can provide more detailed insights into the observed spectro-temporal properties like narrow-band emission-behaviour and downward drifting pattern that has been seen across all repeaters \citep{Hessels2k19,Fonseca2k20,Pleunis2021morpho}.
\par
Recently CHIME reported the discovery of a highly active repeating \src{} \citep{2024ATel16420....1S}. Starting on January 14, 2024, CHIME/FRB detected three bursts from \src{} within a week, and at a DM of 527.7 \ppcc consistent for all the bursts. The CHIME/FRB daily exposure time of only 4 minutes for this source suggested that a high burst rate above the fluence thresholds of $\sim 1$ Jy ms \citep{2024ATel16420....1S}. There were prompt follow-up observations from various telescopes including Parkes, FAST and ours using various frequency bands (300-500 MHz, 550-750 MHz and 1060-1460 MHz) of the upgraded giant metrewave radio telescope \citep[uGMRT;][]{ygupta2017}, which confirmed the high activity of \src{} \citep{2024ATel16430....1U,2024ATel16434....1P,2024ATel16452....1K}.
\par
Subsequently, \src{} was localized to an arc-second precision by the MeerKAT telescope \citep{2024ATel16446....1T}. Following this localization, we focused on low-frequency study of \src{} using band-3 (300-500 MHz) of uGMRT. One of our observations was on 05 March when there was a burst storm as reported by FAST \citep{2024ATel16505....1Z}. Since then, several other detections have been reported from P-band to frequencies up to 6 GHz. \citep{2024ATel16430....1U,2024ATel16432....1O,2024ATel16433....1Z,2024ATel16434....1P,2024ATel16452....1K,2024ATel16494....1P,2024ATel16505....1Z,2024ATel16547....1P,2024ATel16542....1S,2024ATel16446....1T,2024ATel16565....1O,2024ATel16597....1H,2024ATel16599....1J,2024ATel16420....1S}.
\par
Low frequency study of active repeating FRBs provide constraints to various proposed progenitors models for repeating FRBs \citep[e.g.,][]{2021Natur.596..505P}. They also help in studying some of the propagation effects and the underlying emission mechanism \citep{10.1093/mnras/stad3856, 2021ApJ...911L...3P}. Narrow-band behaviour has been reported for repeating FRBs \citep{2021MNRAS.500.2525K} and it has been also been linked to giant pulses from pulsars and magnetars \citep{2021MNRAS.508.1947T,Maan19b}. 
Magnetars appear to be the plausible originators of some of the repeating FRBs, especially since the detection of a millisecond duration burst from the Galactic magnetar SGR~1935+2154 with an energy comparable to the faintest FRBs \citep{2020Natur.587...54C,2020Natur.587...59B}. The low-frequency observations further help in studying the wide-band spectral behaviour of FRBs and probe any potential chromatic evolution of the activity \citep{2021Natur.596..505P,2021ApJ...911L...3P}. 

\par
In this paper, we report detection of 60 bursts at low radio frequencies and %%% 
narrow-band emission from \src{}. We present the results from the observations done in band-4 (550-750 MHz), band-3 (300-500 MHz) and band-5 (1060-1460 MHz) of the upgraded GMRT. We describe the observation and the search procedure in Section~\ref{sec:observations}, the rate function at 400 MHz and extreme narrow-band emission in Section~\ref{sec:results}. Burst properties are detailed in Section~\ref{sec:results}, we discuss the implications of our results for observed properties of the repeating FRBs and their proposed formation channels in Section~\ref{sec:discussions}. Finally, we conclude in Section~\ref{sec:conclusion}.

\begin{table*}[]
\caption{Details of the observations}
\centering
\begin{threeparttable}
\begin{tabular}{l|l|l|l|l|l}\hline
Date        & Frequency       & N$_{antenna}$ & N$_{antenna}$ & Visibliites & T$_{obs}$  \\
            & range           & (IA)$^{*}$    & (PA)$^{\dag}$ & Resolution  &  (min)  \\ \hline
01 Feb 2024 & 550-750 MHz     & 19            & 13            & 0.67s       & 114    \\
08 Feb 2024 & 300-500 MHz     & 18            & 11            & 0.67s       & 104    \\
12 Feb 2024 & 300-500 MHz     & --            & 13            & 0.67s       & 71     \\
13 Feb 2024 & 1060-1460 MHz   & 19            & 5             & 0.67s       & 65     \\
05 Mar 2024 & 300-500 MHz     & --            & 21            & 10.7s       & 55     \\
08 Mar 2024 & 300-500 MHz     & --            & 21            & 10.7s       & 85     \\
24 Mar 2024 & 300-500 MHz     & --            & 22            & 10.7s       & 75     \\
29 May 2024 & 550-750 MHz     & --            & 21            & 10.7s       & 126    \\
15 Jun 2024 & 550-750 MHz     & --            & 23            & 10.7s       & 126    \\ \hline
\end{tabular}

\begin{tablenotes}
\item \xspace $^{*}$ No. of antennae in the incoherent-array mode
\item \xspace $^{\dag}$ No. of antennae in the phased-array mode
\item \xspace T$_{obs}$ is the total on-source time
\end{tablenotes}
\end{threeparttable}
\label{tab:my-table}

\end{table*}

\section{Observations and search procedure}    \label{sec:observations}

Following a possible high level of activity from the repeating \src{} reported by  \citep{2024ATel16420....1S}, we conducted prompt follow-up observations using the uGMRT.
We utilized the wide frequency coverage of uGMRT, spanning 300 MHz to 1460 MHz. Specifically, we observed the source for 114 minutes in band-4 (550–750 MHz) on 01 February 2023, 104 and 71 minutes in band-3 (300–500 MHz) on 08 and 12 February 2023, respectively, and 65 minutes in band-5 (1060–1460 MHz) on 13 February 2023. These observations utilized recording in three different beams: phased-array (PA) beam and the PA spectral voltage (PASV; spectral voltage recording at the Nyquist rate) beam utilizing antennas in compact configuration, and the incoherent array (IA) beam.
We simultaneously recorded the interferometric visibility data with a sampling time of 0.67 seconds to potentially localize the source within a few arcseconds using a burst with adequate signal-to-noise ratio (S/N).
On 12 Feb 2024, we formed four PA beams with different antenna configurations to potentially obtain a localization better than the position known at that time \citep[about 1.5 arcmin uncertainty,][]{2024ATel16420....1S}.
\par
Sensitivity of our above observations was limited as the actual source was about 1.2\,arcmin away from the best estimated position at that time. After \src{} was localized by MeerKAT to RA = 21h27m39.83s and Dec = 04d19'46.02" \citep{2024ATel16446....1T} with about an arcsec precision, we followed it for six more hours at band-3, now utilizing the full sensitivity of the PA beams. Observations were done in three epochs on 05, 08, and 24 March 2024 with the cumulative on-source times of 55, 85, and 75 minutes, respectively. More recently, we also had two observations at band-4, on 29 May 2024 and 15 June 2024, with cumulative on-source times of 126 minutes in each. The summary of all the observations is detailed in Table~\ref{tab:my-table}. In each of these sessions, we recorded the PA beam with 4096 frequency channels and a time resolution of 655 $\mu s$. PASV data were also recorded simultaneously, which were coherently dedispersed and converted to filterbank format with 1024 frequency channels and a time resolution of 81.92 $\mu s$ in the offline processing.
\par
In all the above sessions, we used the following reduction and search procedures. 
The raw data were recorded with 16384, 4096, 4096 frequency channels and with time resolutions of $655.36  \mu s$, $327.28 \mu s$, $163.84 \mu s$ in bands 3, 4 and 5, respectively. For the band-3 observations conducted in March, we converted the PASV data to filterbank format with a sampling time of $81.92 \mu s$ and 1024 frequency channels. All these raw data were passed through a number of RFI excision rounds as follows.
First, RFIClean\footnote{https://github.com/ymaan4/RFIClean} \citep{Maan_2021} is used to mitigate periodic, broadband and spiky RFI as well as convert the data to SIGPROC filterbank format. Subsequently, \texttt{dedispersion} from \textsl{SIGPROC} is used to subband the data to 1024 frequency channels, followed by one more round of RFI mitigation using RFIClean. Finally, we executed \texttt{rfifind} from \textsl{PRESTO} \citep{2011ascl.soft07017R} to compute a mask of further identified RFI-contaminated segments in the data.
\par
We used \texttt{prepdata} from \textsl{PRESTO} to dedisperse over a range of DMs from 522 to 532\,\ppcc, with a step of 0.1\,\ppcc for band-3 and 0.2\,\ppcc for band-4. The dedispersed time-series were then searched for single pulses using \texttt{single\_pulse\_search.py} above S/N of 7 and with a maximum boxcar width of 0.5\,sec. The candidates were clustered in arrival times and DM, and then the candidate corresponding to highest S/N in the individual clusters was verified visually using the waterfall plot prepared using the package \texttt{your} \citep{Aggarwal2020}. 
\par
For the sub-band search, we divided the 200 MHz bandwidth in band-4 as well as band-3 into 7 sub-bands. Each sub-band has a bandwidth of 50 MHz and their the central frequency varies from 325 to 475 MHz for band-3, and 575 to 725 MHz for band-4, in steps of 25 MHz. Thus, there is 50\% overlap between successive sub-bands. The dedispersion for these sub-bands is realized by ignoring the frequency channels not desired for a particular sub-band. The remaining part of the search procedure is same as described earlier. 

\begin{figure*}
    \centering
    \includegraphics[width = \textwidth]{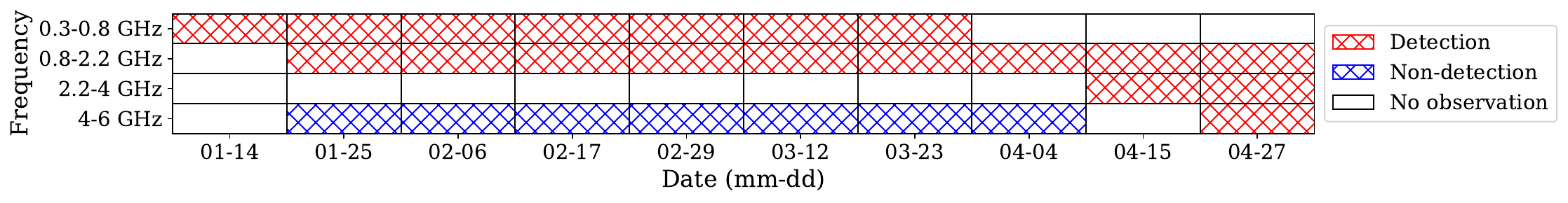}
    \vfill
    \includegraphics[width=\textwidth]{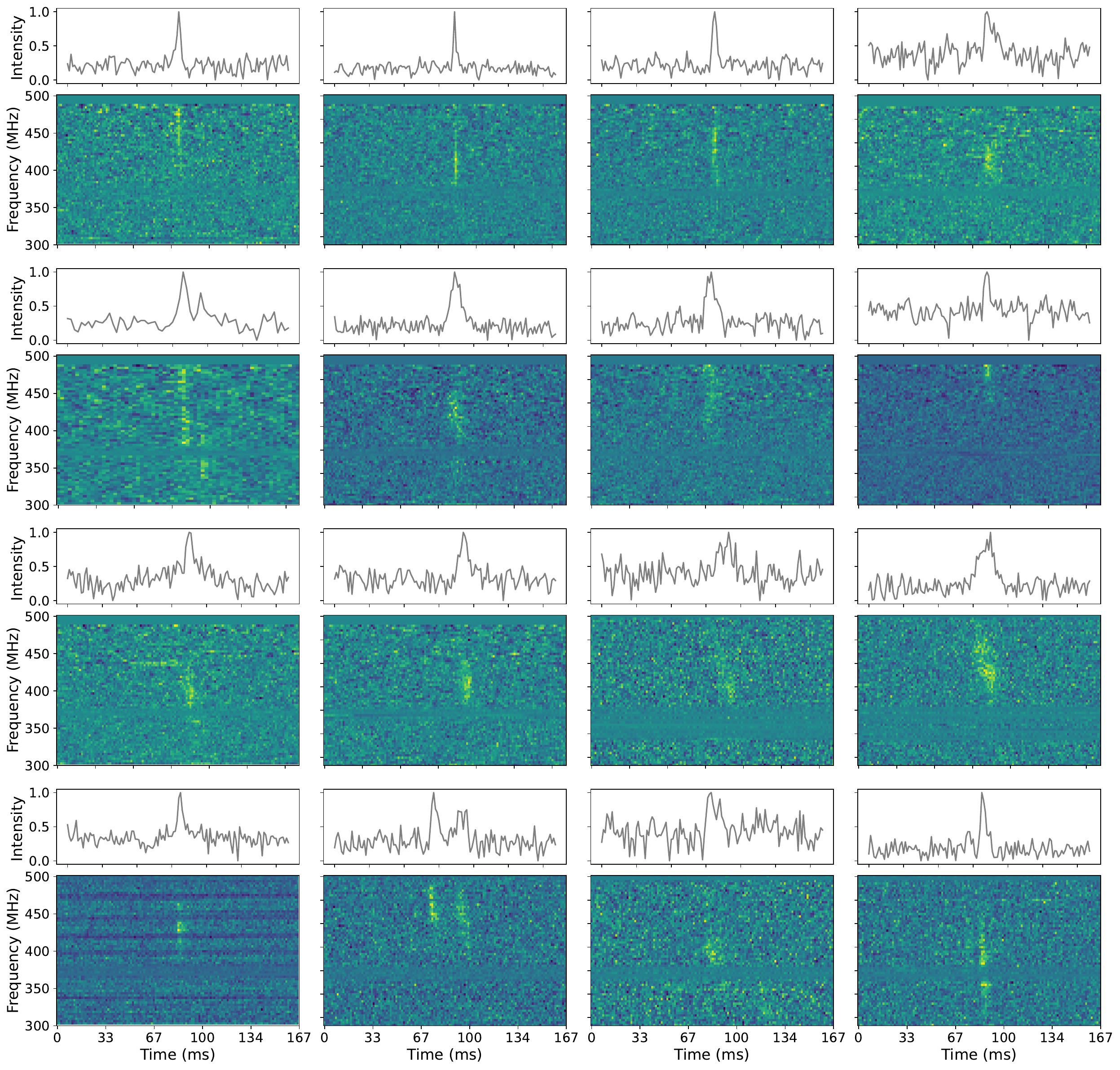}
    \caption{\textit{Top}: A summary of the observations and detection/non-detection of bursts reported at different frequencies, ranging from P-band to C-band, over the duration from 14/01/24 to 09/05/24. The data used here are from various Atels published between 28/01/24 and 09/05/24 \citep{2024ATel16420....1S,2024ATel16430....1U,2024ATel16432....1O,2024ATel16433....1Z,2024ATel16434....1P,2024ATel16446....1T,2024ATel16452....1K,2024ATel16494....1P,2024ATel16505....1Z,2024ATel16542....1S,2024ATel16547....1P,2024ATel16565....1O,2024ATel16597....1H,2024ATel16599....1J,2024ATel16620....1L}. \textit{Bottom:} Dynamic Spectra and time series of some of the bursts detected at band-3 (300-500 MHz).}
    \label{fig:bursts_plots}
\end{figure*}

In the Figure~\ref{fig:bursts_plots} top panel, we show the detection and non-detection of bursts reported at different frequencies in four different ranges: 0.3 to 0.8 GHz, 0.8 to 2.2 GHz, 2.2 to 4 GHz and 4 to 6 GHz, over the duration from 14/01/24 to 09/05/24 roughly every ten days. Here, we use the limited information that is available from various Astronomers Telegrams (ATels) published during this period.

%%%%%%%%%%%%%%%%%%%%%%%%%%%%%%%%%%%%%%%%%%%%%%%%%%%%%%%%%%%%%%

%%%%%%%%%%%%%%%%%%%%%%%%%%%%%%%%%%%%%%%

\section {Burst Analysis and Results} \label{sec:results}

We detected 10 bursts from our band-3 and band-4 observations conducted before the source well localized \citep{2024ATel16446....1T}. In these observations, the pointing center was offset from the actual source position by 1.2\,arcminutes \citep{2024ATel16446....1T}. Of the 10 bursts, 5 were detected in band-4 (550-750 MHz) with an on-source time of 114 minutes, but only in the IA beam indicating that the actual source position was offset by more than the PA beam width (1.1 arcmin). Additionally, we detected 5 bursts in band-3 in the PA beam despite the offset as the PA beam width was adequately larger at these lower frequencies. We corrected the measured fluences for these observations to account for the offset from the actual position. The correction factor is estimated simply as the ratio of the telescope's sensitivity at the pointing center and the actual source position. Consequently, we estimate the burst rate to be 2.6 hr\(^{-1}\) at 650 MHz above the fluence of 0.64 Jy ms, seven days after CHIME detected two bright bursts from \src{} \citep{2024ATel16420....1S}.
\par
From our later observations in March 2024, conducted in band-3 (300-500 MHz) and now using the improved position estimate, we detected significant activity. On 5th March 2024, approximately 3 hours and 15 minutes after a FAST observation wherein a sudden extremely high activity \citep[burst rate of ~500 hr\(^{-1}\) above a fluence of 0.015 Jy ms][]{2024ATel16505....1Z} was noticed, we detected about 27 bursts within 52 minutes, implying a burst rate of 31 hr\(^{-1}\) at a fluence threshold of 0.2 Jy ms. The rate varies significantly over short durations, as we detected only 5 bursts on 8th March 2024, and all within the first ~10 minutes out of nearly 75 minutes of the total on-source time. The rate seemed to have decreased by a factor of ~8 in three days. Later on with observation on 29th May and 15th June we detected 3 very faint bursts in the later session. We do not provide burst analysis with bursts detected in band-4. 
Also, there were several bursts reported at P and L-band after this session \citep{2024ATel16542....1S, 2024ATel16565....1O, 2024ATel16597....1H, 2024ATel16599....1J} implying varying activity level. On 24th March, we detected 15 bursts from an on-source time of 75 minutes. We did not detect any burst from our band-4 session on 29 May 2024.
\par
\begin{figure*}
    \centering
    \includegraphics[width=0.55\textwidth]{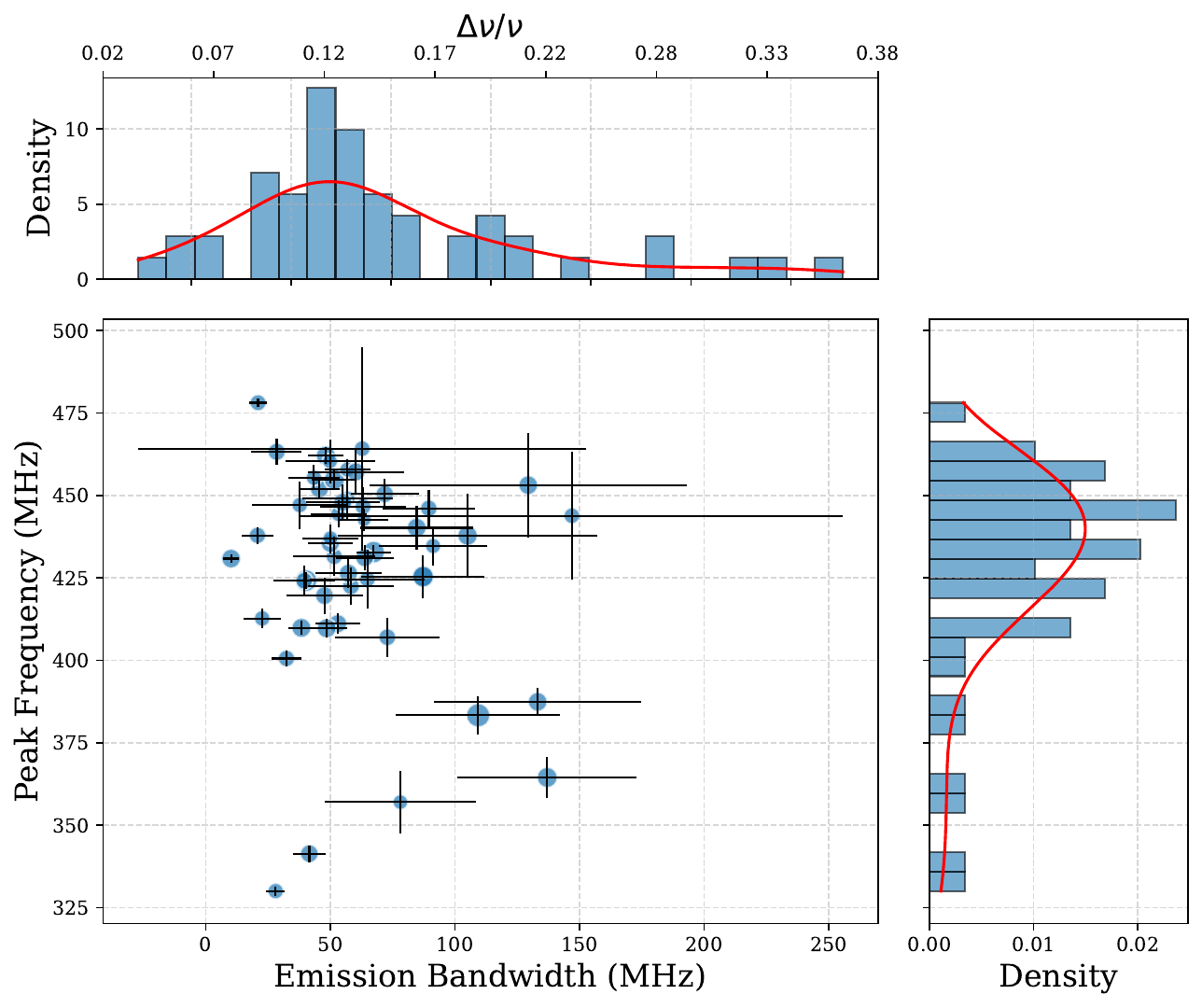}
    \hfill
    \includegraphics[width=0.43\textwidth]{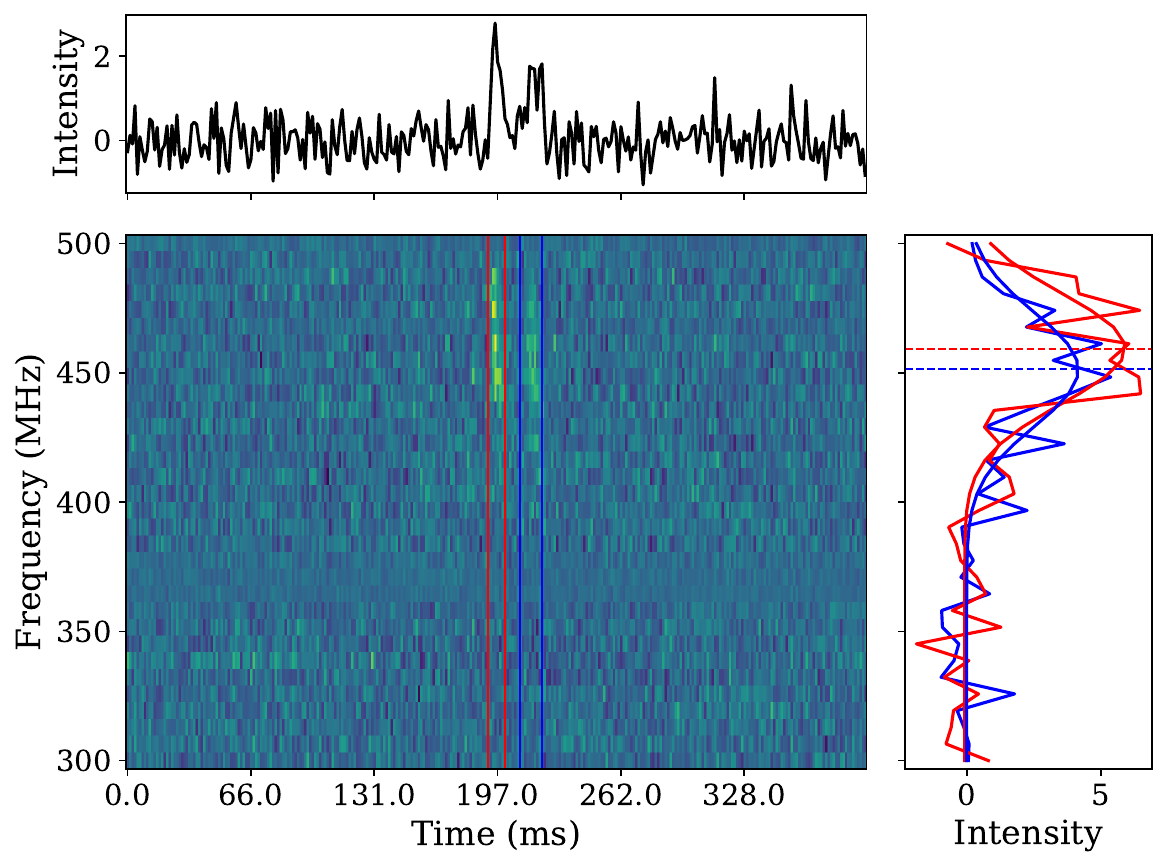}
    \caption{\textit{Left:} Variation of the measured peak frequency with the emission bandwidth. The red line denotes the Gaussian kernel density function for the distribution of the fractional emission bandwidth ($\Delta\nu/\nu$) at top and for the peak frequency at right. \textit{Right:} Example of a burst showing downward drifting pattern and Gaussian fits to the spectra for each component. In the right panel, spectra obtained from averaging along time axis for the regions marked in left panel under each component are shown. The red and green horizontal lines indicate the peak frequency for each of the component.}
    \label{fig:band_limited}
\end{figure*}

In total, we detected 52 bursts in band-3 (300-500 MHz) and 8 bursts in band-4 (550-750 MHz). Some of these bursts exhibit a downward drifting pattern and majority are narrow-band in nature, as shown in Figure~\ref{fig:bursts_plots}. 
Many bursts are primarily centered in the higher part of band-3, i.e., above 400 MHz, as evident from Figure~\ref{fig:band_limited}, however, it is most likely caused by typical more RFI contamination in the central and lower half of band-3.

%%%%%%%%%%%%%%%%%%%%%%%%%%%%%%%%%%%%%%%%%%%%%%%%%%%%%%%%

%%%%%%%%%%%%%%%%%%%%%%%%%%%

\subsection{ Rate Function }  \label{sec:energetics}

\begin{figure}
    \centering
    \includegraphics[scale=0.4]{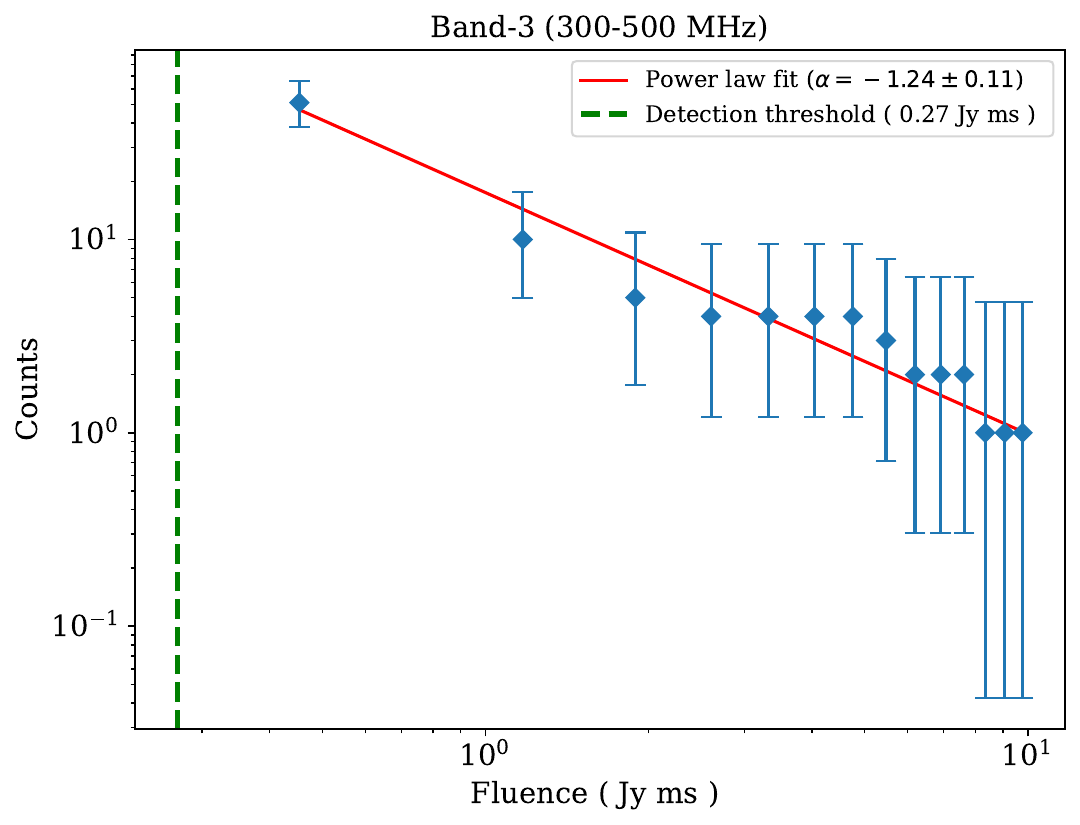}
    \caption{Cumulative distribution of the fluences of the bursts detected at band-3 (300-500 MHz). The green dashed line indicates the fluence threshold for a 2 ms burst detected at a S/N of 10. The red line indicates the fitted power-law.}
    \label{fig:rate_function}
\end{figure}

For each of the bursts, we use the filterbank files and the RFI mask described in Section~\ref{sec:observations}, to prepare a time series dedispersed at the detection DM (i.e., the DM which maximizes the S/N). We extract the time series around the burst's arrival time and normalize the time series in units of the root mean square (RMS) noise measured from off-pulse region. The time series is then calibrated to Jy units using the radiometer equation.
To obtain the fluence for each burst, we summed the calibrated time series around the pulse region. After measuring the fluence for each pulse, we characterize the rate function of \src{} at 400 MHz. Figure~\ref{fig:rate_function} shows the log\,N - log\,S plot. We fitted a power law using the optimization function \texttt{scipy.optimize.minimize}, taking into account Poissonian errors as described by \citep{cash_stat}. All bursts have fluences below 10 Jy ms, suggesting that we are probing the lower end of the energy distribution. However, there have also been bright burst reported with fluences in excess of 10 Jy ms at P-band \citep{2024ATel16565....1O, 2024ATel16547....1P}.
\par
To measure the pulse widths, we used the \texttt{scipy} module \texttt{curve\_fit} to fit a Gaussian profile to each of the burst (component). The pulse widths range from 0.6 ms to 17 ms. The median pulse width of \src{} at 400 MHz is 4.3 ms, which is broader compared to 1.4 ms at 650 MHz\citep{2024arXiv240509749P}. The median for the SNR-optimised DM is 528.1\,\ppcc.However, we do not take into account scatter-broadening while fitting Gaussian function to time profile of bursts.

%%%%%%%%%%%%%%%%%%%%%%%%%%%%%%%%%%%%%%%%%%%%%%%%%%%%%%%%%%%%%%%%%%%%%%%%%%%%%%%%%

\subsection{Emission bandwidths} \label{sec:fractional_bandwidth}

To characterize the emission bandwidth, we first down-sample the data from 1024 to 128 frequency channels and then average along the time axis around the burst region to estimate the average burst spectrum, as illustrated in the right panel of Figure~\ref{fig:band_limited}. The average spectrum is the fitted with a Gaussian function to estimate the peak frequency as well as the emission bandwidth (full width at half maximum).More than 95\% of the bursts exhibit narrow-band emission, as also apparent in Figure~\ref{fig:bursts_plots}.

In Figure~\ref{fig:band_limited}, two major trends are apparent. First, the peak emission frequency seems to decrease (roughly from 480 to 440\,MHz) as the emission bandwidth increases from about 20 to 60\,MHz. It is most likely caused by the band-edge and search sensitivity effects --- near the top band-edge, we might be seeing only parts of the burst emission otherwise continuing outside our observing bandwidth. A similar but inverted trend would be expected at the lower edge of the band, and we see some hints of it. However, several parts of the spectrum in the lower half of the band (300-400\,MHz) are typically contaminated by RFI, which effectively degrade the sensitivity in this part of the band. Consequently, we see this inverted trend effectively starting from roughly around the middle of the band. The distribution of the peak frequencies peaking around 450 MHz is also a consequence of the above band-edge effects. Figure~\ref{fig:band_limited} demonstrates that the emission is narrow-band for nearly all the bursts. The fractional bandwidth distribution peaks around 0.12.

%%%%%%%%%%%%%%%%%%%%%%%%%%%%%%%%%%%%%%%%%%%%%%%%%%%%%%%%%%%%%%%%%%
\subsection{Frequency drift rates} \label{sec:driftrate}

\begin{figure}
    \centering
    \includegraphics[scale = 0.37]{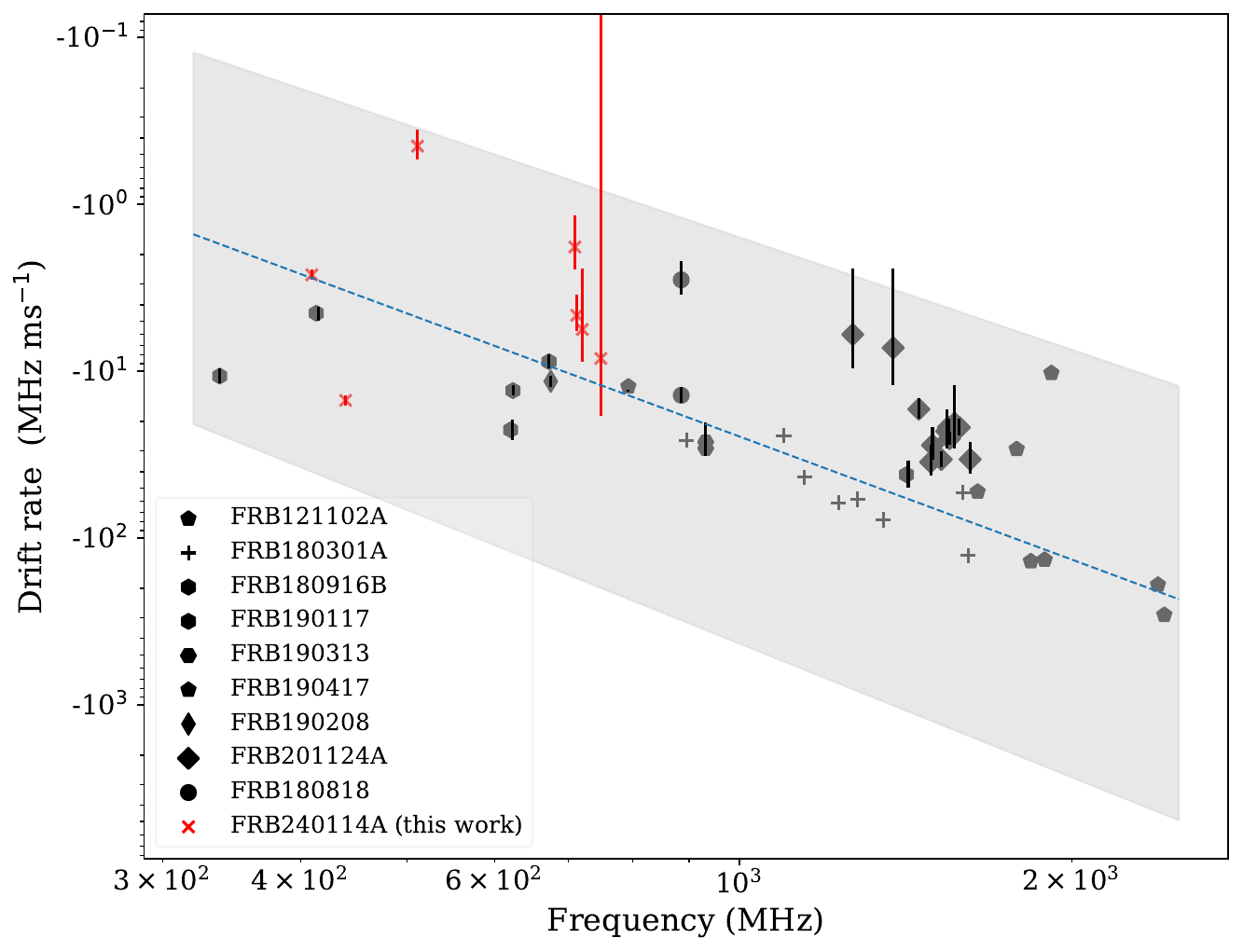}
    \caption{Drift rates as function of frequency, with both the quantities in the rest frame of the respective host galaxy, for \src{} and other repeaters (e.g. FRB121102A, FRB180916B). The dashed blue line and the region shaded around it indicate the power-law fit with its uncertainty, fitted by \citep{Wang2022}. References for measurements on other FRBs: \citet{fast_r67,pravir2023,chime2019b,hessels2019,ines2021,marthi2020}.}
    \label{fig:drift_rates}
\end{figure}

The downward-drifting pattern, is characteristic of repeating FRBs, with different repeating FRBs showing a wide variety of drift rates ranging from $-1$ to $-60$ MHz/ms \citep{2021ApJ...923....1P}. Many of the bursts exhibit this downward drifting with single and multiple sub-bursts. The drift rates are typically estimated by first determining the structure-optimized DM and then exploiting 2-dimensional correlation. Due to the low S/N of our bursts, we could not utilise \texttt{DM\_phase\footnote{\url{https://github.com/danielemichilli/DM_phase}}} \citep{seymour2019} for determining the structure optimized DM. We also could not use \texttt{frbgui\footnote{\url{https://github.com/mef51/frbgui}}} \citep{Chamma_2023} for the same reason. 
So, we use the S/N optimized DM to measure the relative TOAs and the central frequencies of the successive components in multi-component bursts to estimate the drift rate. The peak frequency of the sub-bursts are estimated as described in Section~\ref{sec:fractional_bandwidth}. The relative TOAs are determined by fitting a Gaussian function to each component in the band-averaged time profile. There are three bursts for which we could measure drift rates. One example of a burst for which we measured the drift rate is shown in right panel in Figure~\ref{fig:drift_rates}. The drift rate for this burst is estimated to be $-0.35 \pm 0.07$ MHz/ms. For the other bursts, the drift rate is estimated to be $-11.7 \pm 0.8$ MHz/ms and $-2.07 \pm 0.14$ MHz/ms. These measurements are comparable to the drift rates obtained for other repeaters at similar frequencies, as shown in Figure~\ref{fig:drift_rates}. 
\par
Using the data made publicly available by \citet{2024arXiv240509749P}, we also measure drift rates for 4 bursts detected at band-4 which show well-separated, drifting components. The arrival times of the components are provided by the authors and we estimate the corresponding peak frequencies visually. We liberally assign large uncertainties (3\% - 6\%) to these visual estimates of peak frequencies, which often cover at least half of the total emission bandwidth for the respective components. The drift rates thus estimated are also shown in Figure~\ref{fig:drift_rates}.

%%%%%%%%%%%%%%%%%%%%%%%%%%%%%%%%%%%%%%%%%%%%%%%%%%%%%%%%%%%%%

%%%%%%%%%%%%%%%%%%%%%%%%%%%%%%%%%%%%%%%%%%%%%%%%%%%%%%%%%%%%%%%%%%%%

\section {Imaging analysis and results}

Our observations in February (at band-3 as well as band-4) utilised recording of the interferometric visibilities at 0.67 s integration time with the aim to localize the source to arcsecond precision \citep[the position uncertainty at that time was about 1.5\,arcmin][]{2024ATel16420....1S}. However, these observation did not result in detection of a burst with adequately high S/N for such a localization. Once the source was localized to arcsec precision by \citep{2024ATel16446....1T}, our remaining observations in band-3, used the new position as the pointing center with an integration time of 10.7 s. For band 3 observations, we used 3C48 as the flux calibrator, and for band 4, we used 3C147 and 3C48 in different sessions, and with J2130+050 as the phase calibrator in all the observations.  We used \texttt{CAPTURE\footnote{\url{https://github.com/ruta-k/CAPTURE-CASA6}}} \citep{2021ExA....51...95K} for RFI flagging and calibration of the data.

\begin{figure*}
    \centering
    \includegraphics[width=1\textwidth]{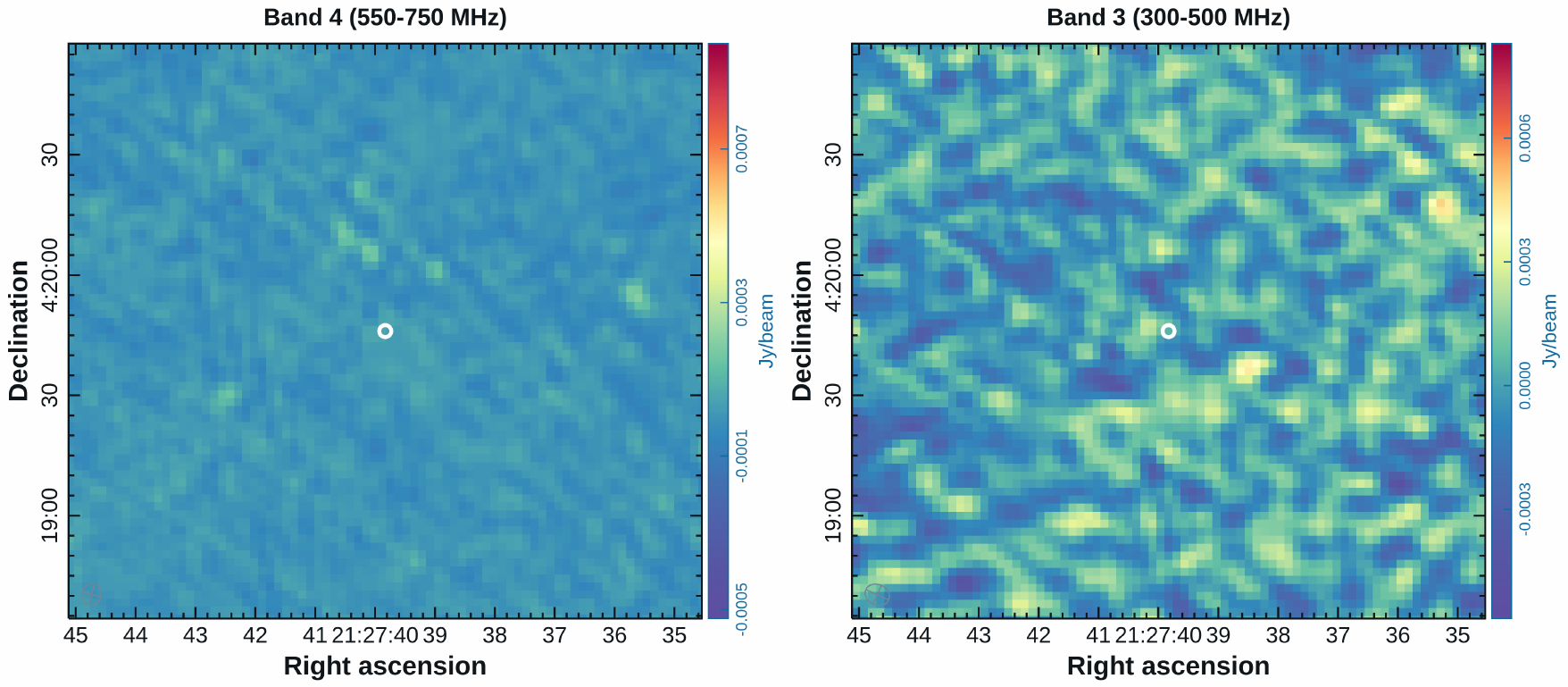}
    \includegraphics[scale = 0.5]{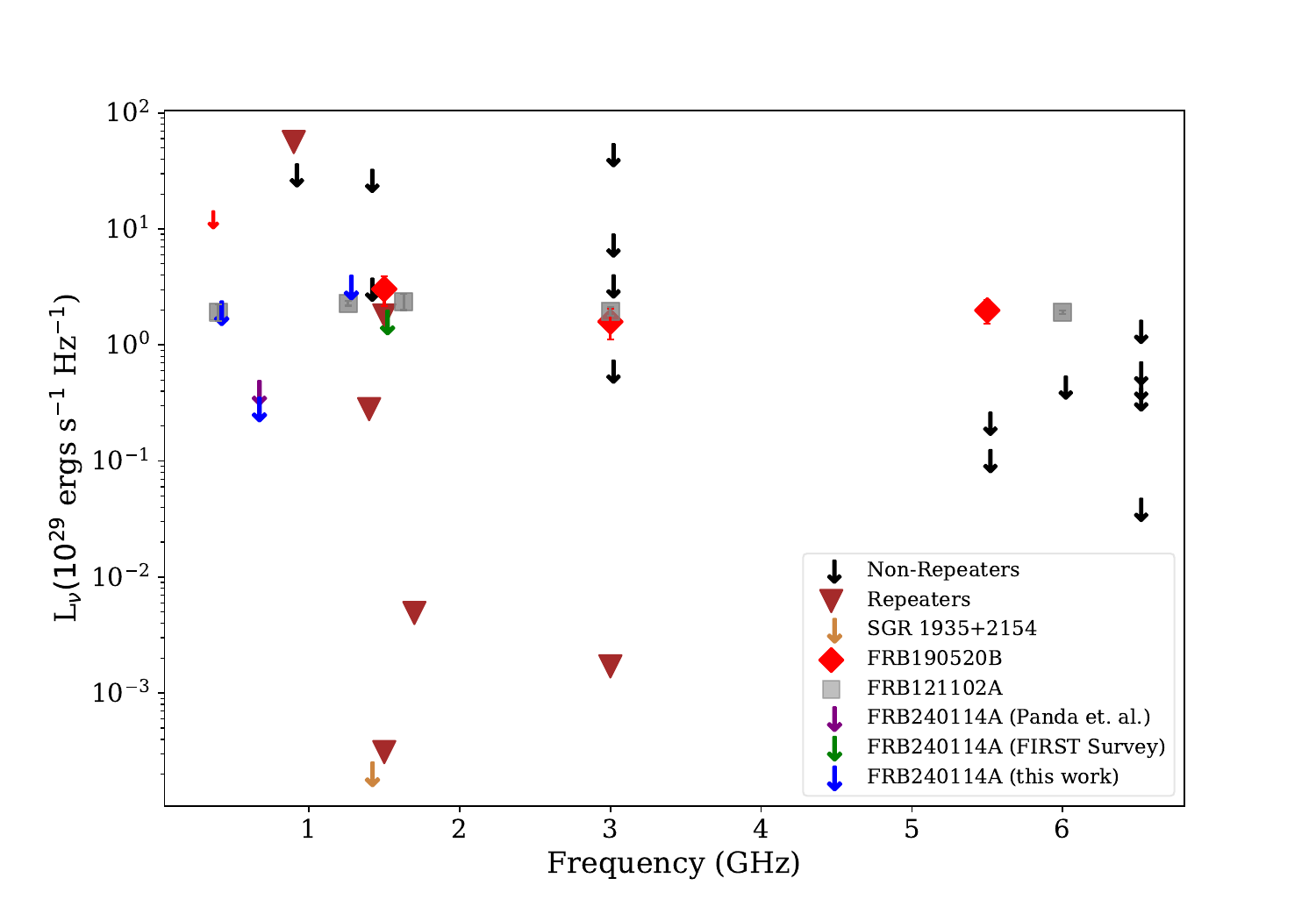}
    \caption{\textit{Top:} Images of the sky regions around the sky-position of \src{}, at 650 MHz and 400 MHz. The MeerKAT localisation of \src{} is marked with a circle at the center of the images. \textit{Bottom:} Comparison of upper limits and measurements of spectral luminosity of PRS, associated with repeating and non-repeating FRBs, and SGR 1935+2154, along with our upper limits for \src{} at different frequencies. References: \citet{2022casey,refId0,Chatterjee2017,zhang2023,niu2022}. For the original references for upper limits on various repeaters, non-repeating FRBs and SGR 1935+2154, please see \citet{2022casey}.} 
    \label{fig:prs_comparison}
\end{figure*}

After calibrating the data, we used CASA task \texttt{tclean} for imaging, with the deconvolution algorithm set to \texttt{MTMFS} and the Briggs weighting scheme (robust = 0). Self-calibration was performed until convergence to an image was achieved. We used \texttt{uGMRTprimarybeam} for primary beam correction for the band-4 image obtained from the 01 Feb 2024 session, since our pointing was slightly offset from the source position. From this observation, we place an upper limit of 140 $\mu$Jy (at 5$\sigma$ level) on any continuum emission associated with the FRB \citep{2024ATel16452....1K}. From our more recent observation on 15 June 2024 we do not detect any continuum emission at the source position, and hence, we place the most stringent 5$\sigma$ upper limit of 89 $\mu$Jy for continuum emission from any associated PRS\footnote{During the review of this paper, \citet{prsatelr147} have suggested the presence of an associated continuum emission of $72\pm14$\,$\mu$Jy at 1.3\,GHz.} or the host galaxy at 650\,MHz \citep[][placed a 5$\sigma$ upper limit of 124 $\mu Jy$ at the same frequency]{2024arXiv240509749P}. Similarly, at band-3, we obtained the final images individually for the datasets from 5, 8, and 24 March observations. In the best image, we do not detect any continuum emission at the source position up to 600 $\mu$Jy (at 5$\sigma$ level) with an on-source time of 78 minutes. We note that there is a bright (325 mJy) source between the FWHM and first null of the uGMRT primary beam. Thus, deconvolution errors limits the RMS noise achievable for the band-3 image despite the longer on-source time. We provide the upper limits for spectral luminosity at 650 MHz and 400 MHz using the redshift of 0.13 \citep{2024ATel16613....1B}. We also obtain image in band-5 with on source time of 65 min but the observations were heavily affected by RFI. Nevertheless, we provide an upper limit of 1 mJy for the PRS at 1.26 GHz using these band-5 data.
Figure~\ref{fig:prs_comparison}\footnote{L$\nu$ is calculated assuming a flat spectrum.} shows the comparison of our upper limits on emission from a persistent radio source (PRS) associated with \src{}, along with that for PRSs associated with FRB121102A and FRB190520B, and upper limits on similar emission from various repeating and non-repeating FRBs, as well as SGR 1935+2154.

% %%%%%%%%%%%%%%%%%%%%%%%%%%%%%%%%%%%%%%%%%%%%%%%%%%%%%%%%%%%%%%%%%%%%%%%%%%%%%%%%%%%%%%%%%%%%%%%%%%%%%%%%%%%%%%%%%%%%%%%%%%%%%%%%%%%%%%%%%%%%%%%%%%%%%%%%%%%%%%%%%%%%%%%%%%%%%%%%%%%%%%%%%%%%%%%%%%%%%%%%%%%%%%%%%%%%%%%%%%%%%%%%%

% 

%%%%%%%%%%%%%%%%%%%%%%%%%%%%%%%%%%%%%%%%%%%%%%%%%%%%%%%%%%%%%%%%%%%%%%%%%%%%%%%%%%%%%%%%%%%%%%%%%%%%%%%%%%%%%%%%%%%%%%%%%%%%%%%%%%%%%%%%%%%%%%%%%%%%%%%%%%%%%%%%%
%%%%%%%%%%%%%%%%%%%%%%%%%%%%%%%%%%%%%%%%%%%%%%%%%%%%%%%%%%%%%%%%%%%%%%%%%%%%%%%%%%%%%%%%%%%%%%%%%%%%%%%%%

\section{Discussion}  \label{sec:discussions}
Our efforts to understand the nature of \src{} included a prompt follow-up in February and March 2024, at band-3 as well as band-4 of uGMRT right after its discovery. From the observation on 1 February at band-4, we detected bursts in the IA beam but not in the PA beam, suggesting that the position reported by CHIME was offset from the actual source position by at least 1.1 arcmin (i.e., half-power beam width), a finding later confirmed by the MeerKAT localization \citep{2024ATel16446....1T}. Subsequent observations in band-3 and band-5 also did not result in bursts adequately bright enough to localize the source. In these early observations, we detected five bursts in Band-3 in the PA beam (but not in the IA beam, as these were all faint bursts), and none at band-5. With these detections, we confirmed the high activity phase of \src{} in the 300-750 MHz frequency range \citep{2024ATel16452....1K}. Using the subsequent observations in March, now with adequately precise position estimate from MeerKAT, we measure significant variations in the activity levels of \src{} which could be related to the underlying emission mechanism or propagation effects. While two repeating FRBs show periodic modulation in their activity \citep{chimeR3,20Rajwade}, a few others show high activity phases (e.g., burst storms) followed by a long quiescence and with no significant periodicity\citep{Lanman_2022}. Currently there is no clear picture what causes such modulations, periodic or otherwise, in the bursting activity.

\subsection{Frequency-dependent activity}
\src{} has been followed up by various telescopes at a variety of observing frequencies covering the range 0.3$-$6\,GHz. We compiled a summary of these observations and burst detections reported by various groups through ATels and show it in Figure~\ref{fig:bursts_plots}. Limited by the details provided in various ATels as well as for the convenience of displaying, we quantize the observing frequency into four ranges: 0.3-0.8, 0.8-2, 2-4, and 4-6\,GHz. Similarly, the time span between the discovery and first week of May is divided into 10 parts. We note that at L-band and lower frequencies (i.e., the first two frequency ranges), bursts from \src{} have been detected regularly. However, at the frequencies higher than 2\,GHz, bursts have been detected only in the later half of April and beginning of May, despite regular observations earlier too. So, there is an indication of a potential chromaticity in the activity, where \src{} is becoming active at higher frequencies at the later epochs. Non-detection of any burst from our latest band-4 observation on 29 May 2024 also indicate towards quenching of the burst activity at lower frequencies.
\par
FRB 20180916B is the only FRB for which a strong evidence of chromaticity in its activity has been seen \citep{ines2021,pluenis2021lofar}. Despite having only a moderate burst rate, the discovery of the chromaticity was aided by the knowledge of the underlying modulation period of the FRB 20180916B. For \src{}, an underlying periodicity, if any, in its activity is yet to be detected. Moreover, the frequency chromaticity in \src{} appears to be opposite to that in FRB 20180916B, i.e., the burst activity at higher frequencies seem to be delayed in \src{} while the opposite is true for FRB 20180916B.
\par
We note that the currently available information on \src{} is not sufficient to establish or rule out above potential chromaticity. For example, bursts at in the 2$-$6\,GHz frequency range have been detected by sensitive telescopes like Effelsberg and Nancay \citep{2024ATel16620....1L,2024ATel16597....1H,2024ATel16599....1J}. However, the observations displayed for the 4$-$6\,GHz range in Figure~\ref{fig:bursts_plots} are mostly by much smaller telescopes and different setups, e.g., with smaller bandwidths \citep{2024ATel16565....1O}. Any lack of sub-band searches might also have contributed to the sensitivity differences. Furthermore, the narrow-band emission of the bursts combined with narrow observing bandwidths might also have been decisive. With more detailed information on various observing setups, it might be possible to conduct simulations to further probe if there is indeed chromaticity in the \src{}'s activity, however, that is beyond the scope of this work. Nevertheless, a systematic, multi-frequency monitoring of the source is needed to better probe this potential chromaticity as well as any underlying periodicity.
\par
Furthermore, \citet{2024ATel16433....1Z} conducted observations on 28 and 29 January 2024, and 01 and 04 February 2024, detecting 38 bursts with a total on-source time of 2 hours. The inferred burst rate at approximately 1.1 GHz is 20 hr\(^{-1}\) above the fluence threshold of around 0.015 Jy ms. Assuming a mean power law index of -1.5, the burst rate would be only 0.072 hr\(^{-1}\) at a fluence threshold of 0.64 Jy ms. However, we estimate the burst rate to be 2.6\,hr$^{-1}$ at 400\,MHz from our first band-3 observation on 08 February 2024, indicating a sharp evolution of burst rate with observing frequency and/or time.

\subsection{Spectro-temporal Properties}
The width distribution (not shown but assessed separately) of the bursts we detected shows a median width of 4.3 ms at 400\,MHz which is larger than the median width of 1.4 ms reported at 650 MHz \citep{2024arXiv240509749P}. In the left panel in Figure~\ref{fig:band_limited}, we see that the peak frequency of majority of the bursts is above 400 MHz, potentially suggesting a preference for the emission frequencies to be above 400 MHz. However, there could also be detection biases, e.g., lower sensitivity in the lower half of band-3 due to presence of RFI and scatter broadening. Most of the bursts in our sample show narrow-band emission. Similar behavior has been observed at 600\,MHz with CHIME and at L-band with FAST \citep{2024ATel16433....1Z,2024ATel16420....1S}. Narrow-band emission has also been observed in the giant pulses of the Crab pulsar and FRB 20190711A \citep{crab_bandlimited,pravir_2021} with similar values of $\Delta\nu/ \nu$ as for \src{}. There is the possibility of bursts being narrow due to emission being below our detection threshold at other frequencies and brightest at frequencies where we detect them, which effectively implies steep modulations in the burst spectra. \citet{cordes2017} shows that caustics can produce strong magnifications ($\leq 10^{2}$) in plasma and present as spectral peaks in the frequency range 0.1-1\,GHz. Plasma lensing has been invoked to explain narrow-band and high-energy bursts for repeating FRBs, e.g., FRB121102A, FRB201124A \citep{chen2024plasma}. \citet{Metzger2019} invokes synchrotron maser emission in ultra-relativistic magnetized shocks to produce the downward drifting pattern. Under certain conditions, it predicts a simultaneous narrow-band emission that is not intrinsic but most likely caused by the propagation effects, e.g. plasma lensing.

\citet{Wang2022} modeled the drift rates of various repeating FRBs as a function of frequency (in the rest frame of the host galaxy) with a power-law. They suggest that the downward drift can be explained through curvature radiation in magnetars originating from different heights. Figure~\ref{fig:drift_rates} shows that the drift rates we measured for \src{} also agree with the relationship modeled by \citet{Wang2022}. In the lower frequency regime, FRB180916B has been the only other FRB with such measurements. Our measurements for \src{} further populate this part of the phase space.

\subsection{Energetics}
Several bursts have now been reported from \src{} that have fluence in excess of 50 Jy ms \citep{2024ATel16432....1O,2024ATel16547....1P,2024ATel16565....1O,2024ATel16420....1S}, however, all the bursts except one we detected have fluences below 10 Jy ms (twice of the CHIME/FRB's typical detection threshold). This might suggest that most of the time the source might be emitting relatively low energy bursts, also seen by \citet{2024ATel16433....1Z}, and high energy bursts might be following a different mean burst rate and power-law statistics. If this is really the case, it would remain to be seen whether the two population of bursts are caused by different emission mechanisms or propagation effects (e.g., plasma lensing). Particularly for FRB20201124A, \citet{kirsten2023connecting} argue that the higher energy burst population cannot be explained by lensing effects and the increased brightness is intrinsic.
\par
For \src{}, we obtain a power-law index of $-1.24 \pm 0.11$ for the bursts we detected at 400MHz. We note that, much like several other works following the same approach, our fluences are estimated from the band-averaged bursts, even for the bursts clearly having narrow bandwidths. Our measured power-law index of $-1.24 \pm 0.11$ consistent with those measured for several other repeaters for which a wide range from relatively flat index of -0.5 to as steep as -4.9 has been reported \citep{kirsten2023connecting}. This index also varies with time and the range of observed energies of the bursts \citep{kirsten2023connecting}.  For \src{}, \citet{2024arXiv240509749P} fit a broken power law for bursts detected at 650\,MHz with index at lower end to be -0.7 which is flatter than what we measure at 400 MHz. While this could suggest a dynamic or frequency-dependent change in the energy distribution, it could also result from incompleteness in search for lower fluence bursts.
\par
A similar power-law index has been measured for several other repeating FRBs as well as the giant pulses from the Crab pulsar and the magnetar XTE~1810$-$197 \citep{maan2019,serylak2009}. However, different measurements have resulted into different power-law indices for the same sources, indicating temporal evolution or a more complex distribution of the bursts.

%%%%%%%%%%%%%%%%%%%%%%%%%%%%%%%%%%%%%%%%%%%%%%%%%%%%%%%%%%%%%%%%%%%%%%%%%%%%%%%%%%%%%%%%%%%%%%%%%%%%%%%%%%%%%%%
\subsection{Persistent Radio Source} 
Persistent radio sources (PRS) have been associated with two repeating FRBs so far, FRB 20190520B and FRB 20121124A \citep{Chatterjee2017,niu2022}. For both FRBs, there is temporal evolution of the associated PRS flux densities \citep{2023MNRAS.525.3626R,Chen_2023}. Figure~\ref{fig:prs_comparison} shows the spectral luminosity of the PRS associated with the above two repeating FRBs at different frequencies, along with the upper limits for a few repeating and one-off FRBs as well as our upper limits on the spectral luminosity of any PRS associated with \src{} at 400 and 650\,MHz. Specially at 650\,MHz, our upper limit is almost an order of magnitude below the spectral luminosity of the two known PRS.
\par
Two models have been invoked to explain the persistent radio emission --- one involves magnetized ion-electron nebula \citep{Margalit_2018} and other involves a hypernebula i.e hyperacrreting compact object \citep{navin2022}. While our limits are quite stringent at lower frequencies, a spectral turn-over around or above 1\,GHz can not be ruled out. In that case, it might still be possible to detect the PRS at higher frequencies. Furthermore, the turn-over is expected to shift to lower radio frequencies with time \citep{Margalit_2018}, and hence the PRS could also become detectable even at 650 MHz at a later epoch. In such a case, our current limits would be useful in constraining the temporal evolution. Overall, a regular sensitive observations at a few different frequencies would be useful to probe any associated PRS.
\par
\citet{2022casey} studied any potential correlations between various burst properties, e.g., the repetition rate, and the associated PRS luminosity. An interesting idea is that the excess DM caused by the host galaxy, \dmhost, might be causally connected to the PRS \citep{2022casey}. For \src{}, \dmhost could be around 330\ppcc \citep{2024ATel16613....1B}, which is larger than the typical contribution from the ISM in the host galaxies. When compared to the \dmhost and repetition rate of FRBs with and without an associated PRS, \src{}'s \dmhost and repetition rate might favor the presence of a PRS. However, \dmhost normalized by the stellar mass of the galaxy could be more informative \citep[e.g., their figure 4]{2022casey}.

%%%%%%%%%%%%%%%%%%%%%%%%%%%%%%%%%%%%%%%%%%%%%%%%%%%%%%%%%%

\section{Conclusions}    \label{sec:conclusion}

We report detection of 60 bursts from \src{} at band-3 (300-500\,MHz) and band-4 (550-750\,MHz) of uGMRT. We also confirm the lower frequency counterpart of the burst storm that was reported at L-band with a rate of 500\,hr$^{-1}$ using FAST on 5th Mar 2024, but with a bursts rate of 31 hr$^{-1} $ above a fluence threshold of 0.22 Jy ms for 2\,ms wide bursts at 400\,MHz. From a detailed analysis of our low-energy burst sample as well as considering the reported detection of \src{} from other telescopes, we conclude the following.
\begin{enumerate}
\item The fluence distribution of the \src{} bursts at 400\,MHz in the range 0.1$-$10\,Jy\,ms is well fit with a single power-law index of -1.24$\pm$0.11.
\item The bursts from \src{} exhibit narrow-band (about 10\% fractional bandwidth) behaviour and downward drifting pattern similar to many other repeating FRBs. Our drift-rate estimates for \src{} are such measurements at the second lowest frequency for any FRB so far. The drift rates are consistent with the power-law fit for such estimates from various FRBs at their rest frame frequencies, and might indicate curvature radiation at different heights as the origin of the observed downward drifting emission.
\item We place the most stringent upper limits at 400 and 650\,MHz on the spectral luminosity of any underlying PRS associated with \src{}. The upper limit at 650\,MHz is nearly an order of magnitude below the luminosity of the two known PRS at this frequency. However, a spectral turnover at higher frequencies can not be ruled out as the cause of the non-detection, and continued monitoring would be helpful.
\item Finally, \src{} shows frequency dependent burst rate. By combining the information available from follow-ups at other telescopes and frequencies, we also show that the burst activity of \src{} is potentially frequency dependent. Our latest observation at band-4 also indicate towards a potential quenching of the activity at low frequencies.
\end{enumerate}

Overall, the similarity of various properties of \src{} to other repeating FRBs suggest a common intrinsic nature of emission. Our results also warrant for a multi-frequency monitoring of \src{} to establish the potential frequency dependent activity and probe a spectral turn-over evolution of any associated PRS.

%%%%%%%%%%%%%%%%%%%%%%%%%%%%%%%%%%%%%%%%%%%%%%%%%%%%%%%%%%%%%%%%%%%%

\section{Software and third party data repository citations} \label{sec:cite}
\facilities{Giant Meterwave Radio Telescope, Khodad, Pune }

\software{astropy \citep{2013A&A...558A..33A},  
          matplotlib \citep{Hunter:2007},
          scipy \citep{2020SciPy-NMeth},
          your \citep{Aggarwal2020},
          RFIClean \citep{Maan_2021},
          PRESTO \citep{2011ascl.soft07017R}
          }

\section*{Acknowledgments}
We would like to thank Ramananda Santra for useful discussions regarding interferometric data analysis. AK would like to thank Arvind Balasubramanian for useful discussions. We would like to thank the Centre Director and the observatory for the prompt time-allocation and scheduling of our observations. YM acknowledges support from the Department of Science and Technology via the Science and Engineering Research Board Startup Research Grant (SRG/2023/002657). We acknowledge the Department of Atomic Energy for funding support, under project 12$-$R\&D$-$TFR$-$5.02$-$0700. GMRT is run by the National Centre for Radio Astrophysics of the Tata Institute of Fundamental Research. 

\vspace{5mm}
\bibliography{sample631}{}
\bibliographystyle{aasjournal}

\end{document}